\documentclass[intlimits,twoside,a4paper]{article}

\usepackage[cp1251]{inputenc}
\usepackage[eqsecnum]{cmpj3}

\issue{2023}{26}{4}{43501}
\doinumber{10.5488/CMP.26.43501}
\title[Poissonian resetting of subdiffusion in a linear potential]%
{Poissonian resetting of subdiffusion in a linear potential}
\author[A. A. Stanislavsky]{A. A. Stanislavsky\orcid{0000-0003-4420-047X}\thanks{Corresponding author: \email{a.a.stanislavsky@rian.kharkov.ua}.}}
\address{
Institute of Radio Astronomy of the National Academy of Sciences of Ukraine, 4 Mystetstv St., 61002 Kharkiv, Ukraine
}

\Keywords{stochastic resetting, stochastic processes, anomalous diffusion, L\'evy flights, nonequilibrium statistical mechanics}

\date{Received March 28, 2023, in final form May 03, 2023}

\begin{document}

\maketitle

\begin{abstract}
Resetting a stochastic process is an important problem describing the evolution of physical, biological and other systems which are continually returned to their certain fixed point. We consider the motion of a subdiffusive particle with a constant drift under Poissonian resetting. In this model the stochastic process is Brownian motion subordinated by an inverse infinitely divisible process (subordinator). Although this approach includes a wide class of subdiffusive system with Poissonian resetting by using different subordinators, each of such systems has a stationary state with the asymmetric Laplace distribution in which the scale and asymmetric parameters depend on the Laplace exponent of the subordinators used. Moreover, the mean time for the particle to reach a target is finite and has a minimum, optimal with respect to the resetting rate. Features of L\'evy motion under this resetting and the effect of a linear potential are discussed.
%
%
\printkeywords
%
\end{abstract}

\section{Introduction}

\begin{figure}[htb]
\centerline{\includegraphics[width=0.65\textwidth]{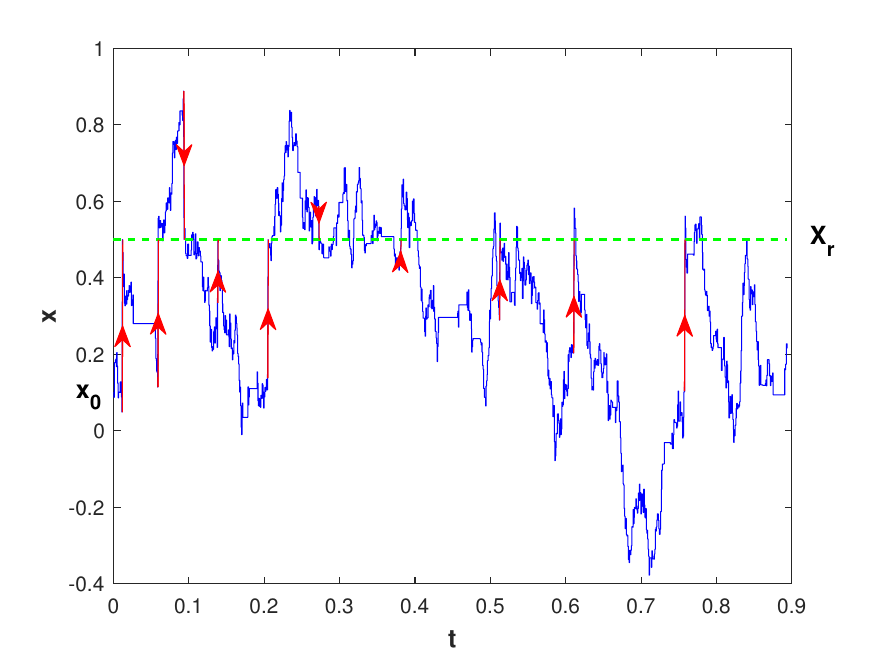}}
\caption{(Colour online) Illustration of subdiffusion ($\alpha=0.95$) with resetting process, manifesting that the particle starts at initial position $x_0$ and resets (red arrows) to position $X_r$ with the rate $r$. Captures by traps are shown by horizontal segments on particle trajectories.}\label{fig:traj}
\end{figure}

Stochastic processes under resetting have attracted considerable attention from the scientific community in recent years (see~\cite{ems20} and its large list of references as well as the special issue~\cite{spec22}). These are explained by a variety of manifestations in nature and in everyday life~\cite{mms20,as21,pga20,r20,tfpsrr20,b20,kgn19}, and the excellent work of Evans and Majumdar~\cite{em11} activated an unprecedented worldwide curiosity to this subject. As has been proven many times over~\cite{b91}, each random search increases the chance of being fruitful, if it uses a strategy in which the searcher goes back to the starting point of his or her search in case of a failure and tries again. The procedure entails not only random walks but also their resetting. On the other hand, the common problem of the searching is to find an optimum research strategy~\cite{bbpmc20}. Moreover, there are different classes of research strategies, and prominent among them is a mixture of local steps and long-range movements~\cite{blm11}. In particular, such strategies can play an exclusive role in the target search of proteins on DNA molecules~\cite{bwh81,cbvm04,bksv09}. The advantage of restarting is also successfully used in many other cases. For example, resetting a stochastic process allows physical systems to reach a stationary state which will be non-equilibrium, due to continually returning to its initial condition~\cite{krbn10}. Complex che\-mi\-cal reactions take advantage of the reset too~\cite{ruk14,dhp14,sddh18}. Thus, the study of stochastic processes subjected to resetting finds new and unexpected applications in science and technology~\cite{tfp08}. While the theory of stochastic resetting demonstrates rather rapid progress in recent years, the experimental side requires new extensions of the topic in various directions that have not been theoretically studied~\cite{gpp20}. This issue is driven by a wide variety of stochastic processes as well as various reset scenarios. In this paper, we pay attention to drift-subdiffusion with resetting. Subdiffusion occurs as often as normal diffusion does everywhere~\cite{szk93,skb02}. It is characterized by a power function in the mean square displacement (MSD) with the exponent less than one. However, subdiffusion can have many faces. The fractional Brownian motion and the subordinated Brownian motion have a similar MSD. The fractional Brownian motion with resetting was considered in the paper~\cite{mo18}. Our study will affect the second case and its generalization, using the infinitely divisible random processes as subordinators. Recently, the normal drift-diffusion with resetting has been comprehensively studied~\cite{rmr19}. In fact, our analysis of the drift-subdiffusion with resetting generalizes the previous results, giving more general relations for non-Gaussian stochastic processes with resetting (figure~\ref{fig:traj}).

The derivation of quantitative analytical results in the theory of stochastic resetting may be based on different approaches. The simplest of them uses Poissonian resetting. In this case, the dynamics includes a stochastic mixture of resetting to the initial position with the rate $r$ (long-range movements) and ordinary diffusion (local steps) with diffusion constant $D$. The probability density governed by the diffusion with resetting can be described with the help of both the master (forward or backward) equation and the renewal equation. All of them give the same results. In the study we use the renewal equation as a simpler method, although the master equation is also used. The paper is organized as follows. Starting with Brownian motion under Poissonian resetting in a linear potential, we emphasize that the stationary distribution takes the asymmetric Laplace form. Moreover, it persists if Brownian motion is subordinated by an inverse infinitely divisible random process. Next, we show that in this case there is also an optimum choice of the resetting rate. Finally, the substitution of Brownian motion for L\'evy flights in a linear potential and under Poissonian resetting gives a stationary state with the asymmetric Linnik distribution.

\begin{figure}[htb]
\centerline{\includegraphics[width=0.65\textwidth]{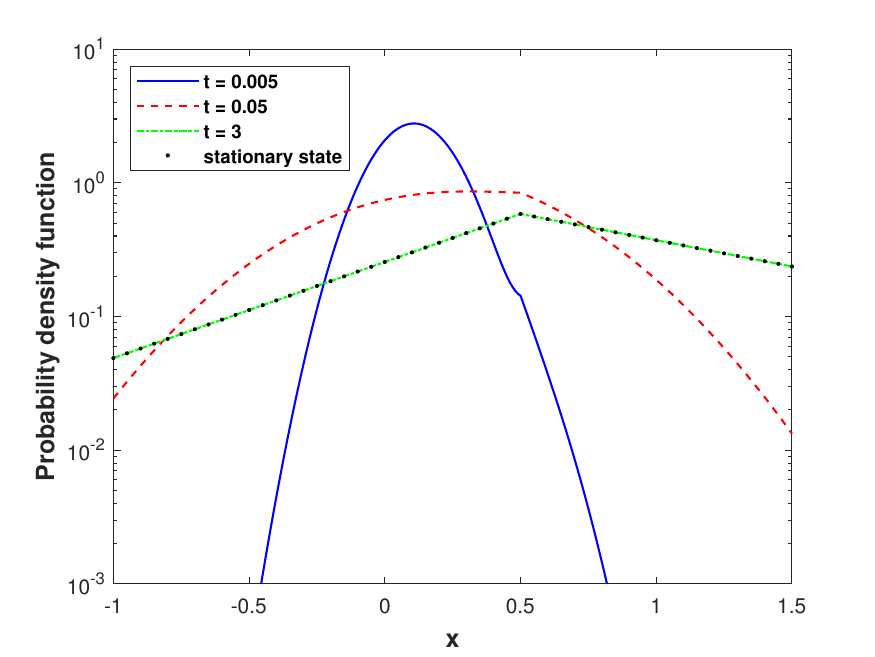}}
\caption{(Colour online) Propagator $p_1(x,t|x_0)$ of ordinary diffusion with Poissonian resetting for $r=3$, $D=2$, $x_0=0.1$, $X_r=0.5$ in potential with $\mu=1.5$, drawn for several instances of time. Starting with the Dirac delta-function at $x_0$ and passing to the subdiffusive PDF, which for $t\to\infty$ becomes the asymmetric Laplace distribution (black dotted line on the panel) with the maximum at $X_r$.}\label{fig:bm}
\end{figure}

\section{Brownian motion in potential}
\label{sec:bm-model}

Brownian motion under resetting has been already considered in potentials and without them~\cite{em11,p15}. It is clear that the former is reduced to the latter, if the potential tends to zero. The interesting result follows from Brownian motion with a constant drift $\mu$ in the positive $x$ direction under Poissonian resetting with the rate $r$. This corresponds to an unbounded linear potential~\cite{ems20}, and the master equation approach to this problem was considered in~\cite{rmr19}. The equation reads 
\begin{equation}
\dfrac{\partial p_1(x,t)}{\partial t}=-\mu\dfrac{\partial p_1(x,t) }{\partial x}
+D\dfrac{\partial^2 p_1(x,t)}{\partial x^{2}}-r p_1(x,t)+r\delta(x-X_r),
\label{eq0}
\end{equation}
with initial condition $p_1(x,0)=\delta(x-x_0)$ as the Dirac delta function, and $X_r$ denotes the position to which the particle returns after resetting. Instead of the master equations we use a simpler way based on renewal equations giving the same solution. For this case, the Green function in the absence of resetting (in other words, the propagator) has the following form
\begin{equation}
G_1(x,t|x_0)=\frac{1}{\sqrt{4\piup Dt}}\exp\left[-\frac{(x-x_0-\mu t)^2}{4Dt}\right],\label{eq1}
\end{equation}
where $x_0$ is the initial position in which $G_1(x,0|x_0)=\delta(x-x_0)$. In the presence of resetting, the probability density function (PDF) $p_1(x,t|x_0)$ is a sum of two terms and is written as
\begin{equation}
p_1(x,t|x_0)=\re^{-rt}G_1(x,t|x_0)+r\int_0^t\re^{-r\tau}\,G_1(x,\tau|X_r)\,\rd\tau.\label{eq2}
\end{equation}
Immediately note that the renewal equation (\ref{eq2}) holds for more general stochastic processes with their propagators, which can be different from this case having $G_1(x,t|x_0)$ as shown in~\cite{sw22}. The evolution of equation~(\ref{eq2}) in time is shown in figure~\ref{fig:bm}. Recall the derivation of equation~(\ref{eq2}) in appendix~\ref{app:deriv}, following from~\cite{ems20}. The stationary state ($t\to\infty$) of equation~(\ref{eq2}) is determined only by the second term which can be found exactly~\cite{ems20}. It acquires the analytical form
\begin{equation}
p_1(x,\infty|X_r)=\frac{r}{\sqrt{\mu^2+4Dr}}\exp\left[{\frac{\mu(x-X_r)}{2D}-\frac{|x-X_r|\sqrt{\mu^2+4Dr}}{2D}}\right].\label{eq3}
\end{equation}
Obviously, the stationary distribution is asymmetric with respect to the resetting position $X_r$ by virtue of different exponential decays in the downstream ($x > X_r$) and upstream ($x < X_r$) directions. Moreover, this is nothing else but the asymmetric Laplace distribution~\cite{kp00,kkp01}. In common notation, the latter reads
\begin{equation}
f(x;m,\lambda,\kappa)=\frac{\lambda}{\kappa+1/\kappa}\begin{cases}
      \re^{(\lambda/\kappa)(x-m)} &\text{for $x<m$}, \\ 
      \re^{-\lambda\kappa(x-m)} &\text{for $x\geqslant m$}.\label{eq3a}
			\end{cases}
\end{equation}
Here, $m$ is the location parameter, $\lambda > 0$ is the scale parameter, and $\kappa>0$ is the asymmetry parameter. As applied to equation~(\ref{eq3}), they depend on the values $X_r$, $r$, $D$ and $\mu$, namely  
\begin{eqnarray}
m &=& X_r,\qquad \lambda=\sqrt{r/D},\nonumber\\
\kappa &=& \left(\sqrt{\mu^2+4Dr}-\mu\right)/\sqrt{4Dr}.\label{eq3b}
\end{eqnarray}
The first two parameters are typical of the ordinary Laplace distribution, whereas the asymmetric para\-me\-ter $\kappa$ leads to the latter distribution, if only $\kappa=1$, i.\,e., $\mu=0$. The result is based on the consideration of Brownian motion as a stochastic process under resetting. Below we show that the asymmetric Laplace distribution is a stationary state for Poissonian resetting of other stochastic processes.

Note that equation~(\ref{eq1}) is a normal variance-mean mixture with respect to both the shift parameter and the scale parameter, but since these parameters are strictly dependent, i.\,e., the position (mean) parameter of the mixed normal law is proportional to its variance, then the mixture (\ref{eq1}) may be characterized by one parameter. That is why the mixture (\ref{eq1}) is called variance-mean~\cite{bnks82,kzk16}. In fact, the function $r\re^{-rt}$ in the integrand of $p_1(x,\infty|X_r)$ is the mixing probability density.

\begin{figure}[htb]
\centerline{\includegraphics[width=0.65\textwidth]{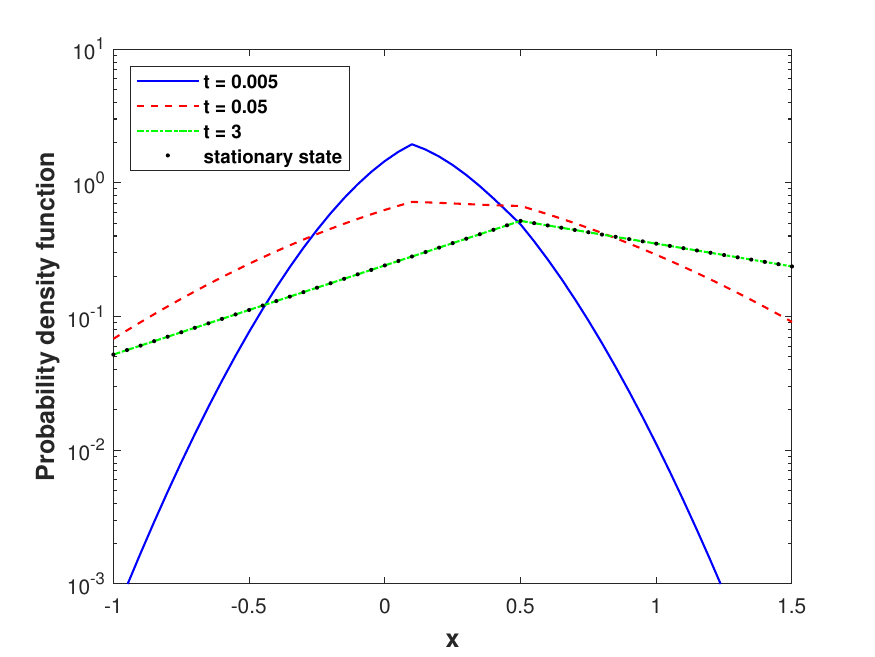}}
\caption{(Colour online) Subdiffusive propagator $p_\alpha(x,t|x_0)$ (obtained from subordination of ordinary diffusion by an inverse $\alpha$-stable process having $\alpha=0.8$) with Poissonian resetting for $r=3$, $D=2$, $x_0=0.1$, $X_r=0.5$ in potential with $\mu=1.5$, drawn for several instances of time. Starting with the Dirac delta-function at $x_0$ and passing to the subdiffusive PDF, which for $t\to\infty$ tends to the asymmetric Laplace distribution (black dotted line on the panel) with the maximum at $X_r$.}\label{fig:sub}
\end{figure}

\section{Ordinary subdiffusion in potential}

Consider the subdiffusion instead of Brownian motion as a stochastic process under Poissonian resetting with the constant rate $r$. Moreover, the subdiffusion includes an unbounded linear potential. In this case, the Green function (also known as the propagator for the subdiffusion equation) is expressed in terms of the subordination integral
\begin{equation}
G_\alpha(x,t|x_0)=\int_0^\infty G_1(x,\xi|x_0)\,g_\alpha(\xi,t)\,\rd\xi,\label{eq4}
\end{equation}
for which the PDF $G_1(x,\xi|x_0)$ describes probabilistic properties of the parent process, whereas the PDF $g_\alpha(\xi,t)$ is related to the directing process which is inverse $\alpha$-stable. The latter has a simple Laplace transform~\cite{mbsb02,s03}
\begin{equation}
\bar g_\alpha(\xi,u)=u^{\alpha-1}\exp\left({-\xi u^\alpha}\right),\label{eq5}
\end{equation}
which we use for our analysis. Let us also remind that the $\alpha$-stable L\'evy process is often considered as a continuous limit of a sequence of nonnegative, independent, identically distributed random variables (obeying an $\alpha$-stable PDF with $0 < \alpha < 1$~\cite{jw94}) representing waiting-time intervals between subsequent jumps of a walker. If such a process is denoted by $U_\alpha(\tau)$, its inverse $\alpha$-stable process is defined as $S_\alpha(t)=\inf\{\tau>0: U_\alpha(\tau)>t\}$. 

Then, the time-dependent renewal equation, accounting for Poissonian resetting, yields
\begin{equation}
p_\alpha(x,t|x_0)=\re^{-rt}G_\alpha(x,t|x_0)+r\int_0^t\re^{-r\tau}\,G_\alpha(x,\tau|X_r)\,\rd\tau,\label{eq6}
\end{equation}
similar to equation~(\ref{eq2}), but with another propagator corresponding to the given case. This equation has a stationary state. An example of evolving to such a state in time is represented in figure~\ref{fig:sub}. For $\alpha=1$, the PDF $g_\alpha(\xi,t)$ becomes the Dirac delta-function (no subordination), and equation~(\ref{eq6}) is transformed into equation~(\ref{eq2}) for Brownian motion. Therefore, the index 1 is used in equation~(\ref{eq2}). As usual, the stationary state is determined by the second term of equation~(\ref{eq6}). Therefore, we consider only this and take:
\begin{equation}
p_\alpha(x,\infty|x_0)=r\int_0^\infty \re^{-r\tau}\,G_\alpha(x,\tau|X_r)\,\rd\tau.\label{eq7}
\end{equation}
From equations~(\ref{eq4}), (\ref{eq5}) and integrating over $\xi$, we find the subduffusive Green function under the potential as the inverse Laplace transform
\begin{equation}
G_\alpha(x,t|X_r)=\frac{1}{2\piup \ri}\int_{\rm Br}\exp\left[{ut+\frac{\mu(x-X_r)}{2D}-\frac{|x-X_r|\sqrt{\mu^2+4Du^\alpha}}{2D}}\right]\,\frac{u^{\alpha-1}\rd u}{\sqrt{\mu^2+4Du^\alpha}},\label{eq8}
\end{equation}
where ${\rm Br}$ is the Bromwich contour, i.\,e., the straight line from $u=\sigma-\ri\infty$ to $u=\sigma+\ri\infty$, where $\sigma$ is chosen so that all the singularities of the integrand are located to the left from the line. In fact, equation~(\ref{eq7}) is the Laplace transform of the function $G_\alpha(x,t|X_r)$, whereas equation~(\ref{eq8}) presents the inverse Laplace transform giving the same function. Thus, the stationary solution of equation~(\ref{eq6}) can be found from the integrand of equation~(\ref{eq8}), calculating the integral
\begin{eqnarray}
p_\alpha(x,\infty|x_0)&=&r\int_0^\infty\delta(s-r)\,\exp\left[{\frac{\mu(x-X_r)}{2D}
-\frac{|x-X_r|\sqrt{\mu^2+4Ds^\alpha}}{2D}}\right]\,\frac{s^{\alpha-1}\rd s}{\sqrt{\mu^2+4Ds^\alpha}}\nonumber\\
&=&\frac{r^\alpha}{\sqrt{\mu^2+4Dr^\alpha}}\,\exp\left[{\frac{\mu(x-X_r)}{2D}
-\frac{|x-X_r|\sqrt{\mu^2+4Dr^\alpha}}{2D}}\right],\label{eq9}
\end{eqnarray}
having many similarities with equation~(\ref{eq3}). In this case, the PDF is also asymmetric about the resetting position $X_r$ because of different exponential decays in the downstream ($x > X_r$) and upstream ($x < X_r$) directions. We again obtain the asymmetric Laplace distribution as a stationary state mentioned above. However, its scale and asymmetry parameters are modified, namely
\begin{eqnarray}
m &=& X_r,\qquad \lambda=\sqrt{r^\alpha/D},\nonumber\\
\kappa &=& \left(\sqrt{\mu^2+4Dr^\alpha}-\mu\right)/\sqrt{4Dr^\alpha}.\label{eq9a}
\end{eqnarray}
The difference between equation~(\ref{eq3b}) and equation~(\ref{eq9a}) is only in the substitution $r\to r^\alpha$, but the PDF form is saved.

\section{Subordination by infinitely divisible processes}
\label{sec:subord}

The next one in line for our consideration is a more general case, when the subordinator is described by an inverse infinitely divisible distribution. Such a distribution has the following Laplace transform~\cite{sw20}
\begin{equation}
\bar g_\Psi(\xi,u)=\frac{\bar\Psi(u)}{u}\re^{-\xi\bar\Psi(u)},\label{eq10}
\end{equation}
where $\bar\Psi(u)$ is the Laplace exponent expressed in terms of Bernstein functions~\cite{ssv10}. In this instance the propagator is
\begin{equation}
G_\Psi(x,t|x_0)=\int_0^\infty G_1(x,\xi|x_0)\,g_\Psi(\xi,t)\,\rd\xi.\label{eq11}
\end{equation}
Using equation~(\ref{eq10}), it is not difficult to find the function $G_\Psi(x,t|x_0)$ as the inverse Laplace transform, namely 
\begin{equation}
G_\Psi(x,t|x_0)=\frac{1}{2\piup \ri}\int_{\rm Br}\exp\left[{ut+\frac{\mu(x-X_r)}{2D}
-\frac{|x-X_r|\sqrt{\mu^2+4D\bar{\Psi}(u)}}{2D}}\right]\frac{\bar{\Psi}(u)}{u\sqrt{\mu^2+4D\bar\Psi(u)}}\rd u.\label{eq12}
\end{equation}
The corresponding Green function $p_\Psi(x,\infty|x_0)$ takes the form of equation~({\ref{eq2}) expressed in terms of the propagator $G_\Psi(x,t|x_0)$. Therefore, we omit it. Obviously, then the second term determines a stationary state of this equation, and this term is nothing else but the Laplace transform for $t\to\infty$. Consequently, the PDF of the stationary state satisfies
\begin{equation}
p_\Psi(x,\infty|x_0)=\frac{\bar{\Psi}(r)}{\sqrt{\mu^2+4D\bar{\Psi}(r)}}\exp\left[{\frac{\mu(x-X_r)}{2D}
-\frac{|x-X_r|\sqrt{\mu^2+4D\bar{\Psi}(r)}}{2D}}\right].\label{eq14}
\end{equation}
Consequently, the brief analysis of the anomalous diffusion with a linear potential leads to the distribution, similar to equations~(\ref{eq3}) and (\ref{eq9}), if $t\to\infty$, in such systems under Poissonian resetting. Then, the parameters, characterizing the asymmetric Laplace distribution, are convenient to represent as
\begin{eqnarray}
m &=& X_r,\quad \lambda=\sqrt{\bar{\Psi}(r)/D},\nonumber\\
\kappa &=& \left(\sqrt{\mu^2+4D\bar{\Psi}(r)}-\mu\right)/\sqrt{4D\bar{\Psi}(r)}.\label{eq14a}
\end{eqnarray}
This result shows that the development affects both scale and asymmetry parameters of the distribution, but its asymmetric Laplace form and peak position do not change.

\begin{figure}[htb]
\centerline{\includegraphics[width=0.65\textwidth]{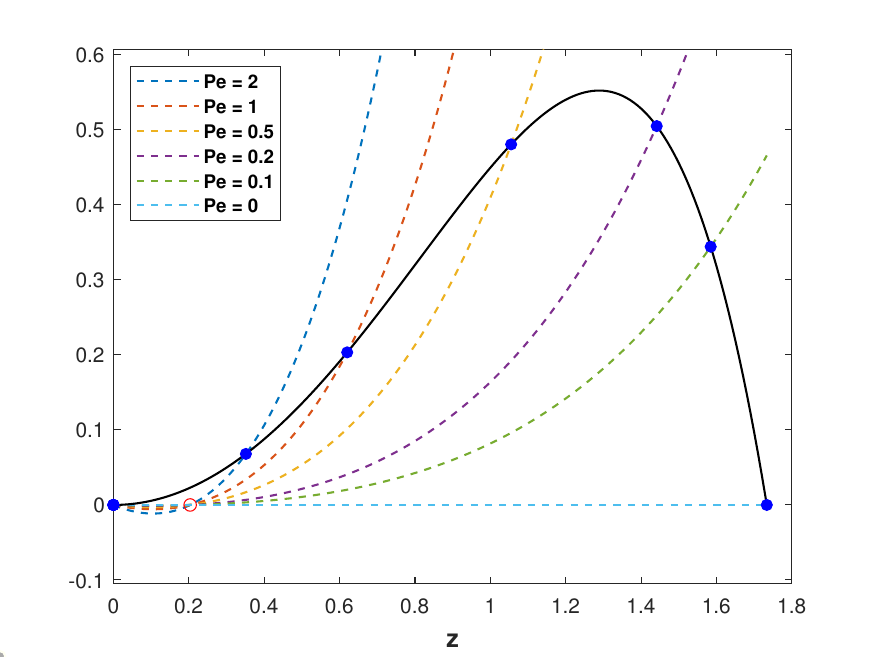}}
\caption{(Colour online) Plots of the left-hand side (colored dotted lines) and right-hand side (black line) of equation~(\ref{eq23}) vs. the reduced variable $z$ from equation~(\ref{eq19}). Blue circles denote solutions to equation~(\ref{eq23}) for different values of the P\'eclet number ($\alpha$ = 0.95).}\label{fig:trans}
\end{figure}

\section{Optimum resetting}

According to~\cite{ems20}, the renewed equation for survival probabilities is written as
\begin{equation}
Q_r(x_0,t)=\re^{-rt}Q_0(x_0,t)+r\int_0^t\re^{-r\tau}\,Q_0(X_r,\tau)\,Q_r(x_0,t-\tau)\,\rd\tau,\label{eq15}
\end{equation}
where  $Q_r(x_0,t)$ denotes $Q_r(x_0,t|X_r)$ shorthandedly, as well as similarly for $Q_0$, to lighten the notations. In equation~(\ref{eq15}) the first term is related to trajectories without resetting, whereas the second term represents trajectories under resetting. After the Laplace transform and when the initial position and resetting position coincide, we have
\begin{equation}
\bar{Q}_r(X_r,s)=\frac{\bar{Q}_0(X_r,r+s)}{1-r\bar{Q}_0(X_r,r+s)}.\label{eq16}
\end{equation}
This is a very general result for Poissonian resetting, also applied to non-Gaussian stochastic processes. Using the Laplace transform of the subdiffusive propagator $G_\Psi(x,t|x_0)$ in the general form (\ref{eq12}), we find the clear expression for $\bar{Q}_0(X_r,r+s)$, using 
\begin{equation}
\bar{Q}_0(x_0,s)=\frac{1-\exp\Big\{-\frac{x_0}{2D}\left[\sqrt{\mu^2+4D\bar{\Psi}(s)}-\mu\right]\Big\}}{s}.\label{eq16a}
\end{equation}
Then, the mean time for the particle to reach a target is $\langle T(X_r)\rangle=\bar{Q}_r(X_r,0)$ expressed in terms of the survival probability $Q_0(X_r,t)$ in the absence of resetting. Thus, the mean time of one-dimensional subordinated diffusion with Poissonian resetting yields
\begin{equation}
\langle T(X_r)\rangle=\frac{1}{r}\left(\re^{z_\Psi}-1\right),\label{eq17}
\end{equation}
where $z_\Psi$ has the following form
 \begin{equation}
z_\Psi = \frac{X_r}{2D}\left[\sqrt{\mu^2+4D\bar{\Psi}(r)}-\mu\right].\label{eq18}
\end{equation}
When $\mu\geqslant0$, the value $z_\Psi$ varies from zero to infinity. However, if $\mu<0$, then $z_\Psi\geqslant {2|\mu|X_r}/{2D}$. This imposes certain conditions on finding the optimum resetting rate. For $r\to 0$ and $r\to\infty$, the mean time $\langle T(X_r)\rangle$ tends to infinity, whereas between them the optimum resetting rate is located.

As an illustrative example, we consider the ordinary subdiffusion having $\bar{\Psi}(s)=s^\alpha$ with $0<\alpha\leqslant 1$. It is convenient to define a reduced variable   
\begin{equation}
z=\frac{X_r}{2D}\left[\sqrt{\mu^2+4Dr^\alpha}-\mu\right],\label{eq19}
\end{equation}
and rewriting the resetting rate $r$ in terms of $z$, we have
\begin{equation}
r=\Biggl[\left(\frac{\mu}{X_r}\right)z+\left(\frac{D}{X_r^2}\right)z^2\Biggr]^{1/\alpha}.\label{eq20}
\end{equation}
\noindent
Substituting equation~(\ref{eq20}) into equation~(\ref{eq17}), the mean time in terms of $z$ is written as
\begin{equation}
\langle T_r\rangle=\frac{X_r^{2/\alpha} z^{1-1/\alpha}}{(X_r\mu+Dz)^{1/\alpha}}\left[\frac{\exp(z)-1}{z}\right].\label{eq21}
\end{equation}
To find the optimum resetting rate, we look for a solution to the equation
\begin{equation}
\frac{\rd \langle  T_r\rangle}{\rd r}=\frac{\alpha X_r^{2/\alpha} z^{1-1/\alpha}}{(X_r\mu+2Dz)(X_r\mu+Dz)^{1/\alpha-1}}
\left[\frac{\rd\langle T_r\rangle}{\rd z}\right]=0.\label{eq22}
\end{equation}
Substituting equation~(\ref{eq21}) into equation~(\ref{eq22}), we get the following transcendental equation 
\begin{equation}
\left[\left(z-\frac{2}{\alpha}+1\right)\exp(z)-1+\frac{2}{\alpha}\right] \mathrm{Pe}
=\left[\left(\frac{1}{\alpha}-\frac{z}{2}\right)\exp(z)-\frac{1}{\alpha}\right]z\,,\label{eq23}
\end{equation}
where $\mathrm{Pe}={\mu X_r}/{2D}$ denotes the P\'eclet number which is the ratio between the rates of advective and diffusive transport~\cite{red01}. For obvious reasons, the value $z=0$ will not be taken into account as a root of equation~(\ref{eq23}). The parameter $\alpha$ characterizes the contribution of random traps in which the particle is frozen in motion for a random time interval from the $\alpha$-stable distribution (figure~\ref{fig:traj}). If $\alpha =1$, no traps exist which is typical of an ordinary diffusion. Another limit case, having $\alpha=0$, manifests the particle confinement forever. In the pure-subdiffusion limit, $\mathrm{Pe}=0$, equation~(\ref{eq23}) is reduced to a simple form
\begin{equation}
1-\exp(-z) = \frac{z\alpha}{2},\label{eq24}
\end{equation}
which has a single root for each value $\alpha$~\cite{kgn19,sw21}. For example, the pure-diffusion case gives $z\simeq1.5936$ mentioned in~\cite{ems20,rmr19}. The tendency of the parameter $\alpha$ from one to zero leads to a monotonous increase in the value of the root to infinity. It should be noted that for $\alpha=1$, the transcendental equation (\ref{eq23}) is simplified to
\begin{equation}
\left[\left(z-1\right)\exp(z)+1\right] \mathrm{Pe}=\left[\left(1-\frac{z}{2}\right)\exp(z)-1\right]z,\label{eq24a}
\end{equation}
obtained in another way for pure diffusion in~\cite{rmr19}. The left-hand side of equation~(\ref{eq23}) can be equal to zero not only because of $\mathrm{Pe}=0$. The point is that the equation 
\begin{equation}
F_1(z,\alpha)=\left[\left(z-\frac{2}{\alpha}+1\right)\exp(z)-1+\frac{2}{\alpha}\right]=0\label{eq25}
\end{equation}
also has a positive root depending on the parameter $\alpha$. Note that the roots (not equal to zero) of equations~(\ref{eq24}), (\ref{eq25}) are different for the same value $\alpha$. The graphic solution of equation~(\ref{eq23}) with the optimum resetting rate is represented in figure~\ref{fig:trans}, plotting  the left-hand side, $F_1(z,\alpha)\,\mathrm{Pe}$ (colored lines), and the right-hand side, $F_2(z,\alpha):=\left\{\left[({1}/{\alpha})-({z}/{2})\right]\exp(z)-({1}/{\alpha})\right\}z$ (black line), of this equation. The solutions, denoted as $z^{\star}>0$, are the $z$ values for which $F_1(z,\alpha)\,\mathrm{Pe}$ and $F_2(z,\alpha)$ intersect. Consequently, for $\mathrm{Pe}>0$, equation~(\ref{eq23}) has a single positive solution $z^{\star}$. Only if $\alpha=1$, for $\mathrm{Pe}<1$, equation~(\ref{eq23}) has one non-trivial positive solution, whereas for $\mathrm{Pe}\geqslant1$ the optimum restart rates are always zero. This effect was called a restart transition~\cite{rmr19}. It means that the restart speeds up the first-passage process for $\mathrm{Pe}<1$, but not for $\mathrm{Pe}\geqslant1$. The feature is due to the fact that the equation $F_1(z,1)=0$ has only a trivial solution at $z=0$ and $\lim_{z\to0}F_2(z,1)=0$, but $F_2(0,1)/F_1(0,1)=1$. Although equation~(\ref{eq23}) has two positive roots for $\mathrm{Pe}<0$, the lesser of them does not satisfy the condition $z\geqslant2|\mathrm{Pe}|$, following from equation~(\ref{eq19}). Therefore, it is just neglected.

\begin{figure}[htb]
\centerline{\includegraphics[width=0.65\textwidth]{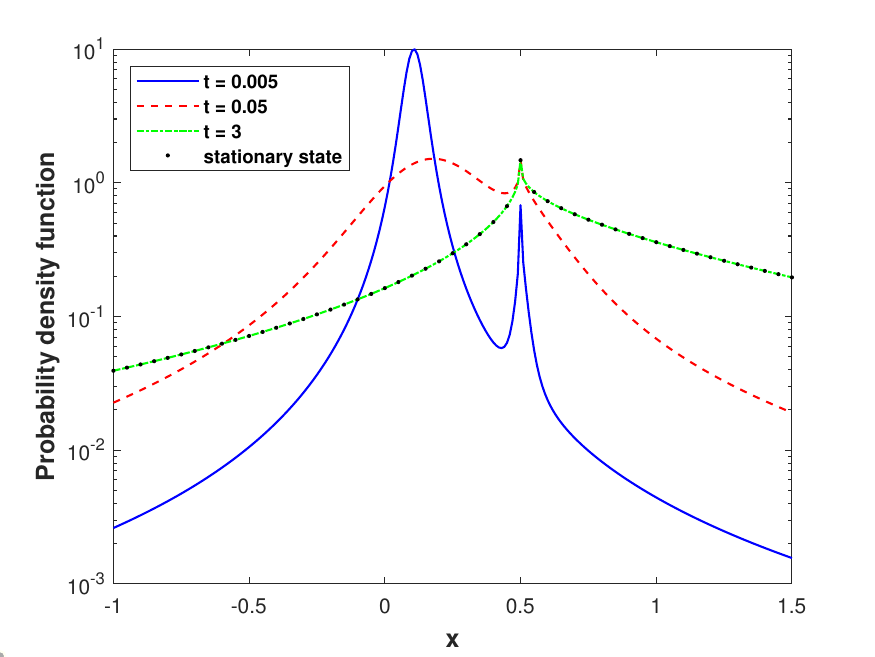}}
\caption{(Colour online) Propagator $p(x,t|x_0)$ of L\'evy motion ($\beta=1.3$) with Poissonian resetting for $r=3$, ${\mathcal D}=2$, $x_0=0.1$, $X_r=0.5$ in potential with $\mu=1.5$, drawn for several instances of time. Starting with the Dirac delta-function at $x_0$ and passing to the L\'evy PDF, which for $t\to\infty$ becomes the asymmetric Linnik distribution, shown by black dotted line on the panel, the maximum of which is located at $X_r$.}\label{fig:levy}
\end{figure}

\section{Asymmetric non-Laplace stationary state}
\label{sec:asym}

The Laplace form of stationary states in equations~(\ref{eq3}), (\ref{eq9}) and (\ref{eq14}) is conditioned by a direct or indirect (through subordination) connection with Brownian motion. If  it is another stochastic process, for example, L\'evy motion, keeping the Poissonian resetting, then the stationary state undergoes changes. Let us consider this case below. L\'evy motion consists of space jumps belonging to the domain of attraction of $\beta$-stable distribution with $0<\beta<2$~\cite{jw94}, and this stochastic process is a a continuous limit of the sum of such jumps. The PDF of L\'evy motion does not have such a clear analytical form as Brownian motion. Therefore, we consider the characteristic function as the Fourier transformation of the Green function, namely
\begin{equation}
\hat{p}(k,t|x_0)=\int_{-\infty}^\infty p(x,t|x_0)\,\re^{\ri kx}\,\rd x.\label{eq26}
\end{equation}
For the $\beta$-stable L\'evy motion, the characteristic exponent is equal to $|k|^\beta$. Applying the resetting equ\-ation~(\ref{eq2}) with the corresponding propagator $G(x,t|x_0)$, it is easy to study this relation in the Fourier space, passing from $x$ to $k$. As a result, the stationary state of the characteristic function is described by the Laplace transform integral
\begin{equation}
\hat{p}(k,\infty|X_r)=r\int_0^\infty \re^{-r\tau}\,\hat{G}(k,\tau|X_r)\,\rd\tau.\label{eq28}
\end{equation}
If $\mu=0$, it is not difficult to get
\begin{equation}
\hat{p}(k,\infty|X_r)=\frac{\re^{\ri kX_r}}{1+{\mathcal D}|k|^\beta/r},\label{eq29}
\end{equation}
where ${\mathcal D}$ is a generalized diffusive constant. This PDF satisfies $p(x,0)=\delta(x)$ or $\hat{p}(k,0)=1$. The term $\re^{\ri kX_r}$ defines the PDF maximum located at $X_r$, and the PDF expression itself is nothing else, but the Linnik distribution~\cite{linnik63}. Linnik distributions are infinitely divisible, and they have an infinite density peak at zero for $0<\beta\leqslant 1$ and a finite peak at zero for $1<\beta\leqslant 2$~\cite{d90}. The L\'evy motion in a constant force field has the same L\'evy distribution as calculated for the free L\'evy motion, but in the Galilei transformed system $x\to x-\mu t$~\cite{jmf99}. Therefore, the general expression $\hat{p}(k,\infty|X_r)$ for any $\mu$ takes the following form
\begin{equation}
\hat{p}(k,\infty|X_r)=\frac{\re^{\ri kX_r}}{1+{\mathcal D}|k|^\beta/r+\ri\mu k/r},\label{eq30}
\end{equation}
describing the asymmetric Linnik distribution~\cite{kp00,kzk16} as a stationary state. The evolution of $p(x,t|x_0)$ to this state is illustrated in figure~\ref{fig:levy}. Here, the value $\mu$ acts as a skew parameter. The asymmetric distribution is obtained so that each branch of it (left-hand and right-hand) turns out to be a copy of the corresponding branches of different symmetric Linnik distributions.

\section{Conclusions}
\label{sec:conclu}

Eventually, we have shown that the linear potential, acting on the subordinated Brownian motion (subdiffusion), leads to an asymmetric form of the Laplace distribution as a stationary state typical of a wide class of subdiffusion under Poissonian resetting. If the potential tends to zero, the stationary PDF takes the ordinary Laplace form~\cite{sw21}. Each asymmetric Laplace PDF is characterized by three parameters: $X_r$ is the location parameter, $\lambda$ is the scale parameter, and $\kappa$ is the asymmetric parameter. The first of them is the same for any stochastic process under resetting, whereas the second one depends on $r$ and $D$. The contribution of $r$ is determined by the Laplace exponent of the stochastic process subjected to resetting, whereas $D$ is not. By changing the rate $r$ (reset protocol) at a constant value $D$ and finding the scaling parameter of the Laplace distribution, one can restore the Laplace exponent of its directing random process. As the stochastic process under resetting is a subordinated Brownian motion, it is not difficult to establish the latter exactly. If the parameter $\kappa\neq 1$, then the subordinated Brownian motion is in a linear potential. Knowing $\kappa$, we can find the value $\mu$ related to the potential. The parameter $\kappa$ is dependent on $r$, $D$ and $\mu$. The L\'evy processes with resetting in potential manifest also a stationary state, but its PDF is a generalization of the asymmetric Laplace distribution in the sense of asymmetric geometrically infinitely divisible PDFs. This analysis also brings us to another interesting feature induced by resetting: the mean first passage time is minimized for an optimum choice of the resetting rate.

\section*{Acknowledgements}

The author kindly acknowledges a partial support of the Polish National Agency for Academic Exchange (NAWA PPN/ULM/2019/1/00087/DEC/1) and thanks Aleksander Weron for fruitful discussions.

\appendix

\section{Derivation of an integral relation}
\label{app:deriv}

Let a particle diffuse under Poissonian resetting. Then, the first renewal equation is written down as
\begin{equation}
p_1(x,t|x_0)=\re^{-rt}G_1(x,t|x_0)
+r\int_{0}^{t}\rd\tau_f\,\re^{-r\tau_f}\,p_1(x,t-\tau_f|X_r),\label{eqA1}
\end{equation}
where the first resetting at time $\tau_f$ has started from $t = 0$. Taking the Laplace transform yields
\begin{equation}
\bar p_1( x,s|x_0) =  \bar G_1( x,r+s| x_0) 
+ r\int_0^\infty \rd t\,\re^{-st}\int_0^t \rd\tau\,\re^{-r\tau}p_1( x,t-\tau| X_r)\,.
\end{equation}
The second term becomes
\begin{equation}
r \int_0^\infty \rd\tau\int_0^\infty \rd t^\prime\,\re^{-st^\prime -(r+s)\tau}\,p_1( x, t^\prime| X_r)
= \frac{r}{r+s}\,\bar p_1( x,s| X_r),
\end{equation}
so that the Laplace transform of the first renewal equation obeys
\begin{equation}
\bar p_1( x,s|x_0) = \bar G_1( x,r+s| x_0)  + \frac{r}{r+s}\,\bar p_1( x,s| X_r)\,.\label{eqA4}
\end{equation}
Taking $x_0=X_r$, the last expression is transformed into the following form
\begin{equation}
\bar G_1( x,r+s|X_r) = \frac{s}{r+s}\,\bar p_1( x,s| X_r) \,.
\end{equation}
Substituting it in (\ref{eqA4}), one obtains
\begin{equation}
\bar p_1( x,s|x_0) = \bar G_1( x,r+s| x_0)  + \frac{r}{s}\,\bar G_1( x,r+s| X_r)\,.
\end{equation}
The inverse Laplace transform gives
\begin{equation}
p_1( x,t|x_0) = \re^{-rt}  G_1( x,t| x_0)
+ r \int_0^t \rd\tau\,\re^{-r\tau}\,G_1( x,\tau| X_r)\,,\label{eqA7}
\end{equation}
which is just the last renewal equation equal to equation~(\ref{eq2}). Using this approach, the equivalence of~(\ref{eqA1}) and (\ref{eqA7}) was proved in~\cite{ems20}.

\ukrainianpart

\title{Пуассонівське скидання субдифузії в лінійному потенціалі}
\author{О. О. Станиславський}
\address{
Радіоастрономічний інститут Національної Академії Наук України, вул. Мистецтв 4, 61002 Харків, Україна 
}
%
%
%

\makeukrtitle

\begin{abstract}
\tolerance=3000%
Скидання стохастичного процесу є важливою проблемою, що описує еволюцію фізичних, біологічних та інших систем, які постійно повертаються до певної фіксованої точки. Розглядається рух субдифузійної час\-тинки з постійним дрейфом при пуассонівському скиданні. У цій моделі стохастичний процес є броунів\-ським рухом, підпорядкованим зворотному нескінченно подільному процесу (субординатору). Хоча цей підхід включає широкий клас субдифузійних систем з пуассонівським скиданням за допомогою різних суб\-ординаторів, кожна з таких систем має стаціонарний стан з асиметричним розподілом Лапласа, в якому масштаб і асиметричні параметри залежать від показника Лапласа використаних субординаторів. Крім того, середній час досягнення частинкою цілі є скінченною величиною і має мінімум, оптимальний щодо швидкості скидання. Обговорюються особливості руху Леві при такому скиданні та вплив лінійного потенціалу на нього.%

\keywords стохастичне скидання, випадкові процеси, аномальна дифузія, польоти Леві, нерівноважна статистична механіка

\end{abstract}

\end{document}